\documentclass[prl,twocolumn,showpacs,amsmath,amssymb,superscriptaddress, longbibliography]{revtex4-1}

\usepackage{graphicx}
\usepackage{dcolumn}
\usepackage{bm}
\usepackage{color}

\usepackage{float,afterpage,wrapfig}
\usepackage[caption=false]{subfig}
\usepackage{multirow}
\usepackage[normalem]{ulem} 
\usepackage{placeins}

\DeclareMathAlphabet{\bi}{OML}{cmm}{b}{it}

\newcommand{\dn}{\downarrow}

\newcommand{\up}{\uparrow}

\begin{document}
\title{
Spin-singlet and spin-triplet pairing correlations in antiferromagnetically coupled Kondo systems}
\author{Haoyu Hu}
\affiliation{Department of Physics and Astronomy, Rice Center for Quantum Materials, 
Rice University,
Houston, Texas, 77005, USA}
\author{Ang Cai}
\affiliation{Department of Physics and Astronomy, Rice Center for Quantum Materials, 
Rice University,
Houston, Texas, 77005, USA}
\author{Lei Chen}
\affiliation{Department of Physics and Astronomy, Rice Center for Quantum Materials, 
Rice University,
Houston, Texas, 77005, USA}
\author{Qimiao Si}
\affiliation{Department of Physics and Astronomy, Rice Center for Quantum Materials,
Rice University,
Houston, Texas, 77005, USA}

\date{\today}
\begin{abstract}
Recent experiments in quantum critical heavy fermion metals have pushed to the fore the question about whether 
antiferromagnetic fluctuations can promote 
both
spin-singlet 
and spin-triplet 
superconductivity.
Here we address the issue through non-perturbative calculations 
in antiferromagnetically correlated Kondo systems.
We identify 
Kondo-destruction quantum critical points in both the SU(2) symmetric and Ising-anisotropic
cluster Bose-Fermi Anderson models.
The spin-singlet pairing correlations are
significantly enhanced near the quantum critical point
in the SU(2) case; however,
with adequate but still realistic degree of
Ising anisotropy, 
the spin-triplet pairing correlations
are competitive.
Our
 results 
demonstrate that 
 spin-flip processes
 strengthen the spin-singlet pairing at the Kondo-destruction quantum critical points,
 and point towards a way for antiferromagnetic correlations to drive spin-triplet pairing.
 Further implications of our findings
 in the broader context of strongly correlated superconductivity are discussed.
\end{abstract}

\maketitle 

Quantum criticality has the potential to be a unifying theme in strongly correlated metals \cite{coleman2005quantum,Pas21.1}.
A quantum critical point (QCP)
arises at the point of continuous transition between two different types of ground states. 
Various types of novel order, including unconventional superconductivity, 
may emerge in its vicinity \cite{broun2008lies}. The materials in which this physics may be relevant include the cuprates \cite{lee2006doping}, organic superconductors \cite{powell2006strong}, 
heavy fermion metals \cite{Kir20.1,steglich2016foundations}, 
and iron-based superconductors \cite{si2016high}. 

Heavy fermion metals represent a particularly important case study for the connection between quantum criticality 
and unconventional superconductivity.
There are a large number -- about 50 -- of them which superconduct. 
In many cases, 
superconductivity develops out of a strange metal
normal state, which is often associated with a QCP at the border of an antiferromagnetic (AF) order. 
Prototype examples are CeCu$_{2}$Si$_{2}$ \cite{Smidman}, 
and
CeRhIn$_5$ \cite{park2006hidden,Kne08.1,Tho12.1}.
Recently, superconductivity has been discovered at ultra-low temperatures
 in the canonical quantum critical heavy fermion metal, YbRh$_2$Si$_2$,
which exhibits AF order and a field-induced QCP
\cite{Smidman}. In addition to the superconducting phase at zero field, which has been 
interpreted as arising from
the intrinsic electronic quantum criticality unmasked by the hyperfine coupling to nuclear spins
\cite{Sch16.1}, 
a new superconducting phase occurs
 near the critical magnetic field \cite{Nguyen2021}.
 The latter appears to have spin-triplet pairing, raising a profound question of how antiferromagnetic correlations 
 can drive spin-triplet pairing in quantum critical Kondo systems.

To make progress, it is necessary to start from a proper description of the quantum criticality in the normal state. 
It has been recognized that heavy fermion quantum criticality can go
beyond the Landau description based on the slow fluctuations of a spin-density-wave (SDW) order
\cite{hertz1976quantum,millis1993effect,moriya2012spin}. Instead, a Kondo-destruction type of 
QCP \cite{si2001locally,coleman2001fermi,Sen04.1}, which involves a partial-Mott (delocalization-localization)
transition and a small-to-large Fermi surface jump, plays an important role \cite{paschen2004hall,gegenwart2007multiple,friedemann2010fermi,shishido2005drastic,park2006hidden,Kne08.1,schroder2000onset}.
Essentially, the Kondo effect is destroyed at the QCP by the dynamical magnetic fluctuations.
As such, the 4$f$ electrons that manifest as the Kondo-driven delocalized composite fermions in the paramagnetic
phase become localized in the AF-ordered phase, leading to a sudden jump of the Fermi surface from large to small.

The questions are, then, two-fold. First, does the Kondo destruction QCP necessarily drive
superconductivity?
 Second, under what conditions does the quantum criticality promote spin-singlet or spin-triplet Cooper pairing?
Theoretically, this problem is challenging because in the quantum critical regime there is no well-defined quasi-particle anywhere on the Fermi surface.
It is thus necessary to search for microscopic models which possess this kind of QCP so that pairing correlations can be studied in a concrete setting.

Both of these questions can be addressed within cluster Bose-Fermi Anderson/Kondo models (BFAM/BFKM). 
These models involve multiple local moments coupled to each other through a Ruderman-Kittel-Kasuya-Yosida (RKKY) interaction 
and simultaneously to both fermionic and bosonic baths. The bosonic bath captures the dynamical magnetic fluctuations 
of a Kondo lattice model, and formally develops from the later through 
a cluster extended dynamical mean field approach (C-EDMFT) \cite{pixley2015cluster}. 
 In the Ising limit (with infinite Ising anisotropy), the cluster BFAM has been studied by the 
continuous-time quantum Monte Carlo (CT-QMC) method
for its 
quantum critical behavior and
 associated pairing correlations \cite{pixley2015pairing}. 
Going beyond the pure Ising case, 
one can expect that the spin-flip processes associated with the transverse component of 
the RKKY interaction will significantly influence the nature of the pairing that develops. However, this is a challenging problem, because the introduction of transverse exchange couplings requires a triple expansion in the CT-QMC approach.
Recently, we have laid the groundwork by introducing an SU(2) CT-QMC method for a single-impurity BFAM that involves a double expansion in the hybridization and transverse Bose-Kondo coupling \cite{cai2019bose} such that it accesses an
extended dynamical range in the quantum critical regime (see also, Ref.\,\onlinecite{otsuki2013spin}.) 

In this Letter, we carry out the first study of an SU(2) symmetric 
cluster BFAM as well as its spin-anisotropic counterpart with finite Ising anisotropy.
To do so, we 
develop the aforementioned triple-expansion scheme within the CT-QMC approach.
We identify a Kondo-destruction QCP 
in the SU(2) symmetric model. Near the QCP, we find that spin-singlet pairing correlations are significantly enhanced, 
more strongly than its Ising-anisotropic counterpart.
At the same time, the spin-triplet pairing correlations are suppressed, which is to be contrasted with
what happens in the realistically Ising-anisotropic case. We understand both features from the physics that spin-flip processes 
strengthen the spin-singlet pairing at the Kondo-destruction QCPs.

\emph{Model and solution methods:}
The two-impurity SU(2) BFAM Hamiltonian is
\begin{eqnarray}
H&=&\sum_{i,\sigma} \epsilon_{d} d_{i\sigma}^{\dagger} d_{i\sigma} + \sum_{i} U n_{i\uparrow} n_{i\downarrow}  
+I\sum_{\alpha}S^{\alpha}_{1} S^{\alpha}_{2}
\nonumber \\
&+&\sum_{{\bf k},\sigma} \epsilon_{\bf k} c_{{\bf k}\sigma}^{\dagger}c_{{\bf k}\sigma}
+\sum_{i,{\bf k},\sigma} \left( V e^{i {\bf k \cdot r_{i}}} d_{i\sigma}^{\dagger} c_{{\bf k}\sigma} + h.c. \right) \nonumber \\ 
&+& \sum_{p,\alpha} \omega_{p} {\phi^{\alpha}_{p}}^{\dagger} \phi^{\alpha}_{p} + g \sum_{p,\alpha}( S_{1}^{\alpha}-S_{2}^{\alpha} )  ({ \phi_{p}^{\alpha}}^{\dagger}
+\phi_{p}^{\alpha}) \, .
\label{eq:su2-model}
\end{eqnarray}
The first line describes the interactions among the local degrees of freedom: $d_{i\sigma}^{\dagger}$ creates an electron on site $i=1,2$ with spin $\sigma=\uparrow,\downarrow$, $\epsilon_{d}$ is the onsite energy, $U$ is the coulomb repulsion and $I$ is the
RKKY interaction between the two impurities. $S_{i}^{\alpha}=\frac{1}{2} \sum_{\mu,\nu} d^{\dagger}_{i\mu} \sigma^{\alpha}_{\mu\nu} d_{i\nu} $ 
is the spin operator and $\sigma^{\alpha}_{\mu\nu} $ are the Pauli matrices.
The second line describes the fermionic bath and its coupling with the impurities: $c^{\dagger}_{{\bf k} \sigma}$ creates a conduction electron with wave vector ${\bf k}$, spin $\sigma$ and energy $\epsilon_{\bf k}$, which hybridizes with the local $d$ electron through the matrix element $V$.
The third line describes the 3-component vector bosonic bath and its coupling to the anti-ferro component of the two local moments through coupling constant $g$.
$\alpha$ is the spin index and it is summed over $x,y$ and $z$ such that the Hamiltonian is SU(2) invariant.
We will also consider the Ising-anisotropic version of the model (see below).

For the fermionic bath, we consider a flat density of states 
$\rho_{F}(\epsilon) = \sum_{\bf k} \delta (\epsilon - \epsilon_{\bf k} ) = \rho_{0} \Theta( |D-\epsilon|)$, with a hybridization function $\Gamma(\epsilon) = \Gamma_{0} \Theta ( | D - \epsilon| )$ and $\Gamma_{0}=\pi \rho_{0} V^{2}$. In addition, the impurity location ${\bf r_{1}}$ and ${\bf r_{2}}$ are set to be infinitely separated to avoid double counting of the RKKY interactions.
The bosonic bath takes a sub-ohmic density of states specified by
$\rho_{B}(\omega) = \sum_{p} \delta(\omega - \omega_{p} ) = K_{0} \omega^{s} e ^{-\omega/ \Lambda} \Theta(\omega)$.
The prefactor $\rho_{0}$ and $K_{0}$ are determined by the normalization conditions: $\int_{-D}^{D}\rho_{F}(\epsilon) d\epsilon=1$ and $\int_{0}^{\infty} \rho_{B}(\omega) d\omega =1$.
The BFAM can be solved by a recently developed sign-problem-free 
CT-QMC method.
The triple expansion method builds on prior work that has dealt with various 
double expansions  \cite{steiner2015double,otsuki2013spin,cai2019bose}. 
Further details of the method are described in the Supplementary Material.
In the calculation, we set $D=1$, $\Lambda=1$, and preserve particle-hole symmetry by choosing $U=-2 \epsilon_{d} = 0.1$. In addition, we take a sub-ohmic bosonic bath with exponent $s=0.6$ and fix the hybridization parameter $\Gamma_{0}$ at $\Gamma_{0}=0.5$. The RKKY interactions $I$ and bosonic coupling $g$ are tuning parameters to explore phase digram and search for QCPs.

\begin{figure}[t]
\captionsetup[subfigure]{labelformat=empty}
  \centering
 	\mbox{\includegraphics[width=0.8\columnwidth]{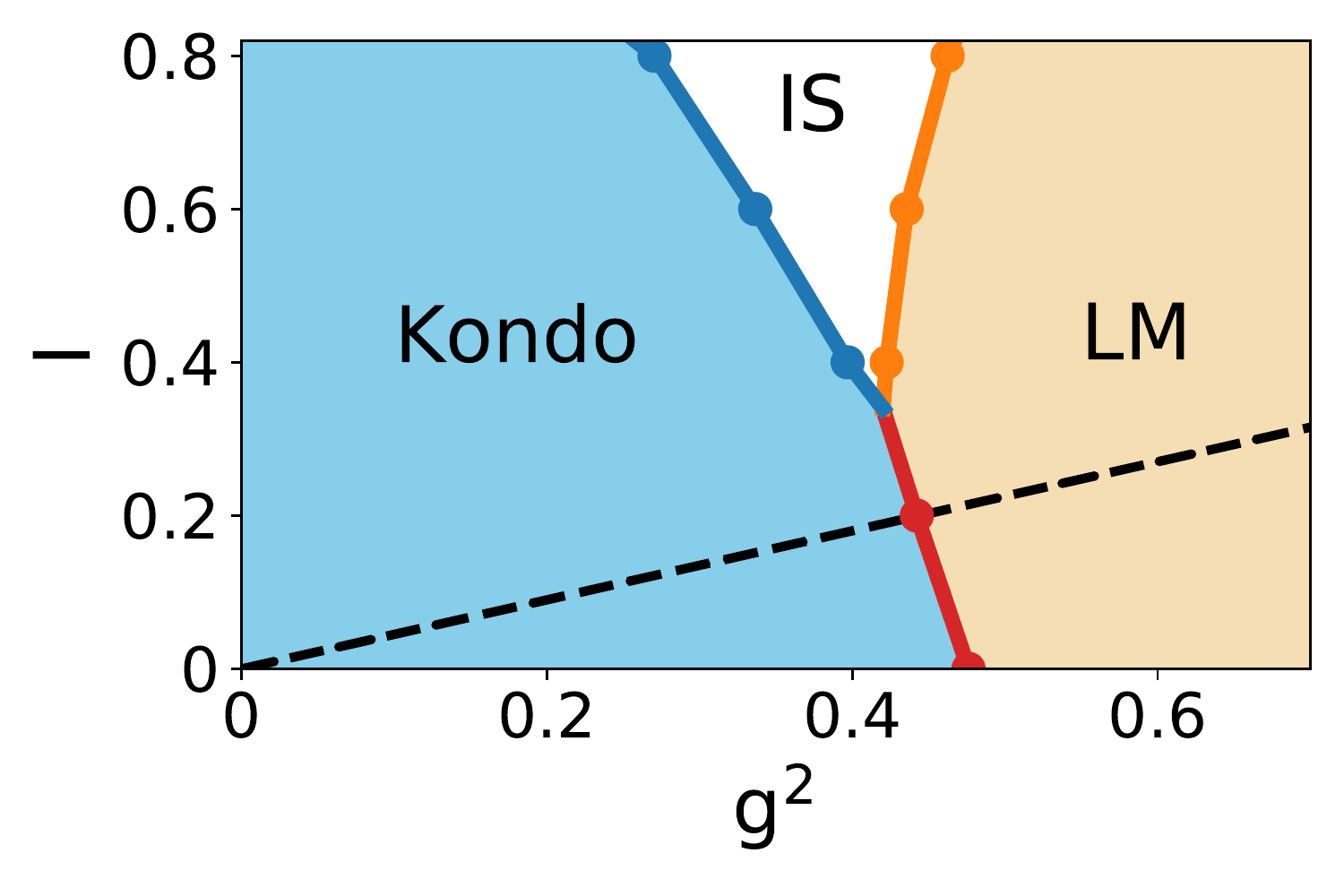}}
\caption{Phase diagram for the two-impurity SU(2) Bose-Fermi Anderson model determined via CT-QMC in the $I$ vs $g^{2}$ plane at fixed hybridization parameter $\Gamma_{0}=0.5$. Blue circles represent QCPs from Kondo screened phase to impurity singlet (IS) phase governed by the critical point at $g=0$. Orange circles represent QCPs from IS phase to antiferromagnetic local moment (LM) phase. Red circles represent QCPs from 
the Kondo screened phase to the LM phase, whose critical behavior is governed by the Kondo-destruction fixed point at $I=0$. Dashes line represent the path $(g^{2}, I)=(2.22\lambda,\lambda)$ 
taken to study the pairing correlation, which
 mimics the trajectory in self-consistent C-EDMFT calculations.}
\label{fig:1}
\end{figure}

\emph{Quantum critical properties:}
Fig.\,\ref{fig:1} shows the $T=0$ phase diagram of SU(2) symmetric BFAM in the $g^{2}-I$ parameter space
obtained from CT-QMC. In the absence of the bosonic bath, the model reduces to the two impurity Anderson model \cite{jones1987study}, which has a single QCP from Kondo screened phase at small RKKY interaction to an impurity singlet (IS) phase at large RKKY interaction. 
Turning on the bosonic coupling $g$, we find that the transition persists with the same critical behavior but at a reduced critical value of $I$. 
At large bosonic coupling, there will be a second transition from the IS phase to an antiferromagnetic local moment (LM) phase.
For $I$ smaller than $0.3$, the IS phase disappears and we have a direct transition between a Kondo screened phase and the LM phase.
 This turns out to be a Kondo destruction QCP that has not been studied before.

\begin{figure}[t]
\captionsetup[subfigure]{labelformat=empty}
  \centering
 	\mbox{\includegraphics[width=1\columnwidth]{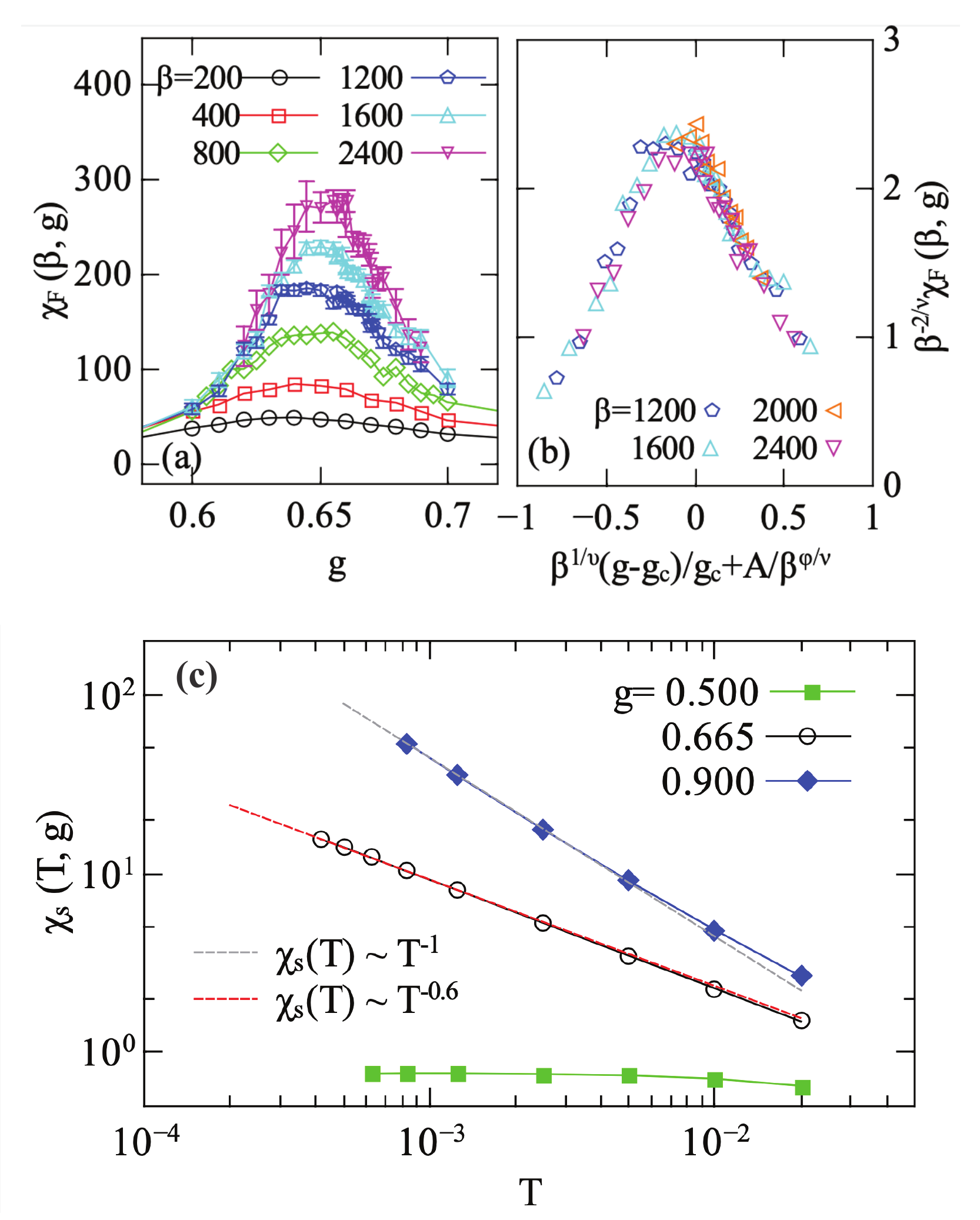}}
\caption{(a) Fidelity susceptibility $\chi_{F}$ vs. bosonic coupling $g$, showing divergence near the critical coupling $g_{c}=0.665(5)$. (b) Finite size scaling of $\chi_{F}$ according to Eq.(\ref{eq:chif}), with  $\nu^{-1}=0.31$, $g_{c}=0.66$.
(c) Staggered spin susceptibility $\chi_{s}(T)$ in the Kondo screened phase (green squares), at the critical point (black circles), and in the LM phase (blue diamonds). The red and grey dashed lines are fits to $\chi_{s}(T)$ at the critical point $g=g_{c}$ and in the LM phase, respectively.
}
\label{fig:3}
\end{figure}

Given its ``beyond-Landau" nature,
we characterize the Kondo-destruction QCP with fidelity susceptibility that directly evaluates the changes of wavefunctions and detects quantum phase transition 
 without introducing any order parameters \cite{zanardi2006ground}.
 The fidelity susceptibility $\chi_F$ is defined as the second derivative of fidelity's logarithm, where the fidelity measures the distance between the ground-state wavefunctions at two different values of parameters $V,V+\delta V$. This definition can be extended to finite temperature: $\chi_{F}=\int_{0}^{\beta/2} \left( \langle {\cal T_{\tau}}  H_{V} (\tau) H_{V} \rangle -\langle H_{V} \rangle^{2} \right)  \tau d\tau$, where $H_V$ includes all the terms in the Hamiltonian that are proportional to the hybridization strength $V$ \cite{albuquerque2010,wang2015fidelity}. Due to the sharp changes of ground-state wavefunctions near the QCP, we observe a singular behaviors of $\chi_F$:
\begin{eqnarray}
\chi_{F}(\beta,g)/ \beta &=& \beta^{2/\nu} \tilde{\chi} \left( \beta^{1/\nu}(g-g_{c})/g_{c} + A/\beta^{\phi/\nu} \right) \, ,
\label{eq:chif}
\end{eqnarray}
with $g_c=0.66(2)$ and $\nu^{-1}=0.31(6)$. The critical value and critical exponent are obtained by fitting the numerical data as shown in Fig.\,\ref{fig:3} (a), (b). The scaling behavior is consistent with a continuous transition and, moreover, suggests that $V$ is a relevant perturbation at QCP. Thus, the transition is indeed a Kondo destruction QCP.
Besides, we study the Binder cumulants and correlation length of the spin correlation function 
(Supplementary Material).
The results are consistent
with the conclusions drawn above from the fidelity susceptibility.

To further characterize the QCP,
we analyze the static spin susceptibility, defined as $\chi_{s}(T)=\int_{0}^{\beta} \chi_{s}(\tau) d\tau$.
As seen in Fig.\,\ref{fig:3} (c), $\chi_{s}(T)$ diverges as $\chi_{s}(T)\sim T^{-x}$ with $x=0.60(1)$ at the QCP. The result is consistent with the relation $x=s$, suggesting that the critical behavior in the spin channel is completely determined by the properties of the bosonic bath. 
Note that this is one of the necessary conditions for the Kondo-destruction QCP to be realized in the lattice case  \cite{si2001locally}.

\begin{figure}[t!]
\captionsetup[subfigure]{labelformat=empty}
  \centering
 	\mbox{\includegraphics[width=0.8\columnwidth]{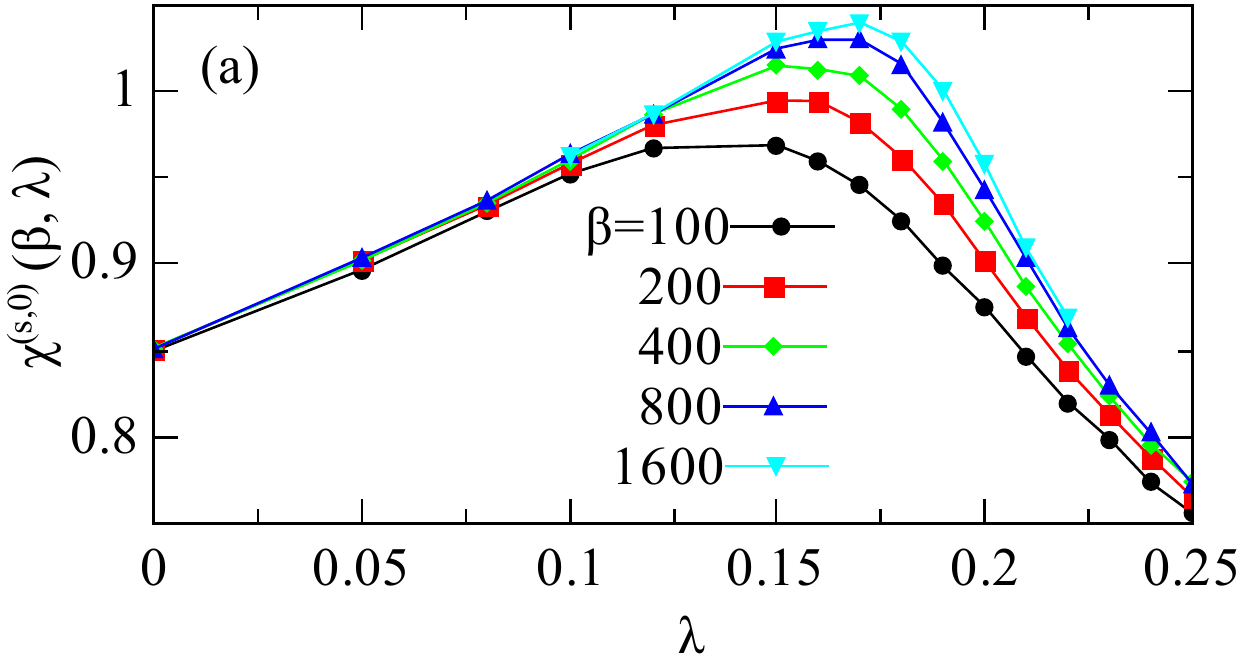}}
  	\mbox{\includegraphics[width=0.8\columnwidth]{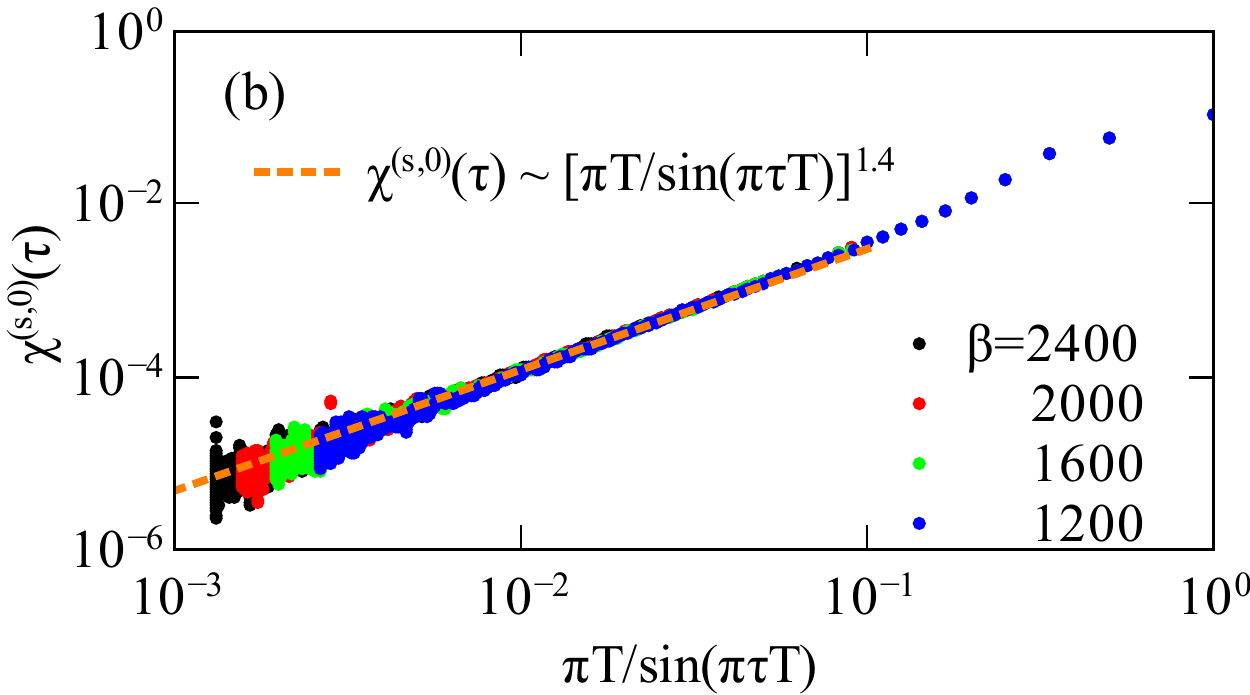}}
\caption{(a) Static singlet pairing susceptibility $\chi^{(s,0)}(T)$ along the cut $\lambda$: $(g^{2}, I)=(2.22\lambda,\lambda)$ as shown in Fig.\,\ref{fig:1}. 
The Kondo destruction QCP is located at $\lambda_c=0.2$. 
(b) Imaginary time pairing correlation function $\chi^{(s,0)}(\tau)$ at the QCP, consistent with a $1/\tau^{2-\eta}$ decay with $\eta=0.6$. 
The dashed orange line is a fit according to $\chi^{(s,0)}(\tau) \sim \left(\pi T / sin(\pi\tau T) \right)^{1.4}$. }
\label{fig:5}
\end{figure} 

\emph{Pairing susceptibilities:}
Within our 2-site construction we can study four different types of inter-site pairing, specified by the following pairing operators: $\Delta_{(t,\pm 1)}^{\dagger} =\Delta^{\dagger}_{\up\up /\dn\dn} $, $\Delta_{(t,0)}^{\dagger} = \left( \Delta^{\dagger}_{\uparrow\downarrow} + \Delta^{\dagger}_{\downarrow\uparrow}  \right) /\sqrt{2} $, and $\Delta_{(s,0)}^{\dagger}  =  \left( \Delta^{\dagger}_{\uparrow\downarrow} - \Delta^{\dagger}_{\downarrow\uparrow}  \right) /\sqrt{2} $, with $\Delta_{\sigma \sigma^{\prime}} ^{\dagger} = d_{1 \sigma} ^{\dagger} d_{2\sigma^{\prime}} ^{\dagger}$, $\{\sigma,\sigma^{\prime}\}  \in \{ \uparrow,\downarrow \}$. They are, respectively, the pairing operators in the triplet channel with $m^{z}_{tot}=\pm 1$, $m^{z}_{tot}=0$ and the singlet channel with $m^{z}_{tot}=0$, where $m^{z}_{tot}$ is the $z$ component of the total spin of the cooper pair.
Accordingly we can define the dynamic and static pairing correlation functions respectively:
$\chi^{(\alpha,m)}(\tau) = \langle{\cal T_{\tau}} \Delta_{(\alpha,m)}^{\dagger} (\tau) \Delta_{(\alpha,m)} \rangle$, $\chi^{(\alpha,m)}(T)= \int_{0}^{\beta} \chi^{(\alpha,m)}(\tau) d\tau$.
The details of the calculational procedure is given in the Supplementary Material.

We study the behavior of the pairing susceptibility following a particular cut $(g^{2},I)=(2.22\lambda,\lambda)$ illustrated in 
Fig.\,\ref{fig:1}. This cut mimics the trajectory of the C-EDMFT self-consistent calculation, and 
will meet the line of Kondo destruction QCPs at $\lambda=\lambda_c=0.2$. 
 As shown in Fig.\,\ref{fig:5}(a), the pairing susceptibility in the singlet channel $\chi^{(s,0)}(T)$ increases as 
 we move from the Kondo screened phase to the QCP, and is peaked slightly before the transition. 
 There is also a large enhancement near the QCP as we lower the temperature: $\chi^{(s,0)}(\tau)$ acquires 
 a large anomalous dimension at the critical fixed point. In Fig.\,\ref{fig:5}(b) we plot $\chi^{(s,0)}(\tau)$ against the conformal form 
 $\chi^{(s,0)}(\tau)\sim \left(\pi T / \sin(\pi \tau T) \right)^{2-\eta}$. We observe a collapse across different temperatures, 
 which suggests that $\chi^{(s,0)}(\tau)\sim 1/\tau^{2-\eta}$, with an anomalous exponent $\eta \simeq 0.6$. 
At the same time, we find that the pairing susceptibility in the triplet channel is strongly suppressed and does not acquire any anomalous dimension.

\begin{figure}[t!]
\captionsetup[subfigure]{labelformat=empty}
  \centering
 	\mbox{\includegraphics[width=1\columnwidth]{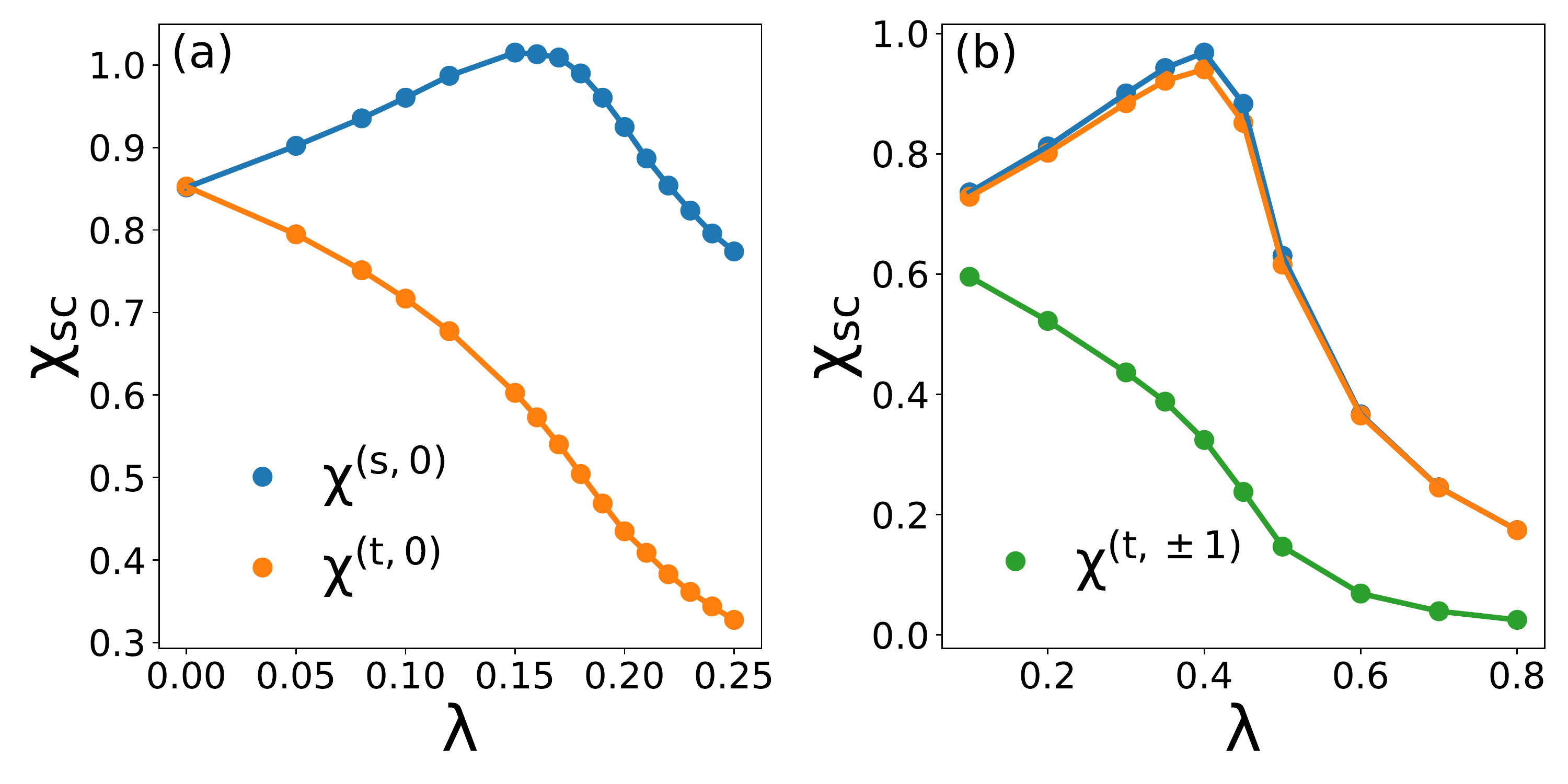}}
 	\caption{Static pairing susceptibilities of the SU(2) symmetric model (a) and the Ising anistropic model (b) at $\beta=400$. 
	(a) SU(2) symmetric model, where the trajectory is the same as in 
	 Fig.\,\ref{fig:1}, with $\lambda_c=0.2$. The degenerate pairing susceptibilities in three triplet channels, $\chi^{(t,0)},\chi^{(t,\pm 1)}$, are 
	 suppressed near the QCP. (b) The Ising anisotropic model, with anisotropy $J_p/J_z=0.2$. We take a cut along 
	 $\lambda:(g,I)=(\lambda,2.688 \lambda)$, with
	 $\lambda_c\sim 0.45$.
	The pairing susceptibility in the triplet $S^z=0$ channel, $\chi^{(t,0)}$, is similarly enhanced near the QCP 
	as for the singlet channel $\chi^{(s,0)}$.}
\label{fig:sc}
\end{figure}

\emph{SU(2) {\it vs.} Ising-anisotropic model}
For the SU(2) model, Fig.\,\ref{fig:sc}(a) displays the evolution of the pairing susceptibilities across the QCP.
It clearly
shows the enhancement of  the spin-singlet pairing correlation near the QCP, and the concomitant 
suppression of the spin-triplet pairing correlation in all the three (degenerate)
 channels.

We now consider an Ising-anisotropic model. The extreme Ising-anisotropic model of Ref.\,\onlinecite{pixley2015pairing}
corresponds to having only the Ising component for both the bosonic-Kondo coupling and RKKY interaction.
More realistically, we consider the model whose effective RKKY interaction is written in an anisotropic form:
\begin{eqnarray}
J_zS^z_1S^z_2 +J_p(S^x_1S^x_2+S^y_1S^y_2 ) \, .
\label{eq:RKKY-anisotropic}
\end{eqnarray}
Here $J_z$ and $J_p$ are respectively the longitudinal and spin-flip RKKY exchange interactions.
The details of the model are described in the Supplementary Material. 

We consider the parameters such that the Ising anisotropy near the QCP has a realistic
 $J_p/J_z \approx 0.2 $.
 Fig.\,\ref{fig:sc}(b) shows that the pairing correlation in the $S_z=0$ triplet channel is enhanced near the QCP, along with the spin-singlet pairing correlation. (By contrast, the pairing correlations
in the $S_z=\pm 1$ triplet channels are suppressed.) 
In the case of extreme Ising anisotropy, with $J_p=0$, the two dominant pairing susceptibilities are exactly equal to each other. Here, the pairing susceptibility in the 
$S_z=0$ triplet channel is almost equal to its counterpart in the spin-singlet channel. 
An analysis of the Landau free-energy functional shows
that, a Zeeman coupling will further enhance the $S_z=0$ spin-triplet channel over the spin-singlet channel,
thereby making the spin-triplet pairing dominate \cite{Hu21.2x}.

\emph{Discussion and Summary:} 
Our findings shed light on the recent experiment in YbRh$_2$Si$_2$ near its field-driven QCP \cite{Nguyen2021}.
YbRh$_2$Si$_2$ has antiferromagnetic order at the ambient conditions. It has a strong $xy$-anisotropy in the spin space, which is turned to Ising-anisotropic by the applied magnetic field.
Our results for the Ising-anisotropic cluster BFAM with an AF RKKY interaction suggest that the $S_z=0$ 
spin-triplet pairing channel will be competitive against the spin-singlet 
pairing near the Kondo-destruction QCP, and can be further turned into the dominant pairing channel by the 
Zeeman coupling of the critical magnetic field. This provides a natural understanding of the surprising experimental results.

By contrast, when the spin anisotropy is relatively weak, our results derived from the SU(2) 
symmetric cluster BFAM suggests that the spin-singlet pairing dominates. This result lays the groundwork 
for the understanding of the superconductivity that has been experimentally observed in a host of antiferromagnetically correlated heavy fermion
metals, such as 
CeRhIn$_5$ and CeCu$_2$Si$_2$.

Our results also shed light on other classes of strongly correlated metals. For example, UTe$_2$ has a strong Ising anisotropy 
in its spin response \cite{Ran2019}.
Recent inelastic 
neutron scattering experiments \cite{Duan21.1x} show that both the normal and superconducting states are antiferromagnetically correlated; 
this raises the exciting possibility that, here too,
antiferromagnetic correlations are primarily responsible for the potential spin-triplet pairing.

Going beyond the heavy-fermion context, the electronic delocalization-localization effect has been implicated by 
experiments in both the hole-doped cuprates \cite{Bad16.1}
  and organic charge-transfer salts \cite{Oik15.1}. Recently, based on analyses of related dynamical equations, 
  it has been suggested that beyond-Landau quantum criticality that bears similarities to the Kondo-destruction 
  QCP operates in doped Mott insulators \cite{Cho21.1x}. These calculations have stayed at the spin-rotational invariant level. 
  Our results suggest that in these models too, it would be instructive to analyze the role of spin anisotropy that may arise in doping Mott insulators such as Sr$_2$IrO$_4$.

  To summarize, we have studied a cluster Bose-Fermi Anderson model as an effective Kondo system for beyond-Landau quantum criticality. In both the SU(2) symmetric and Ising anisotropic models, we identify 
  Kondo-destruction quantum critical points. The spin-singlet pairing correlations are more strongly enhanced near the QCP in the SU(2) symmetric case than its Ising-anisotropic counterpart.
  This reveals that spin-flip processes strengthen the spin-singlet pairing.
   In accordance with this insight, we find the $S_z=0$ spin-triplet pairing channel to be competitive 
   in the realistically Ising-anisotropic version of the model.
The results provide a natural understanding of the recent experiments on the superconductivity near the field-driven quantum critical point 
of the prominent heavy fermion system YbRh$_2$Si$_2$, in a way that incorporates the salient quantum critical characteristics in its normal state.
Finally, our findings bring out new insights pertinent
to the unconventional superconductivity in a broad range of strongly correlated metals.

\emph{Acknowledgements.} 
We thank P. C. Dai, K. Ingersent, S. Paschen, J. H. Pixley, F. Steglich, J. D. Thompson and J.-X. Zhu for useful discussions.
The work was in part supported by the National Science Foundation under Grant No.\ DMR-1920740 (H.H. and L.C.),
the Air Force Office of Scientific Research under Grant No.\ FA9550-21-1-0356 (Q.S.),
the Robert A. Welch Foundation Grant No.\ C-1411 (A.C.),
 the Data Analysis and Visualization 
Cyberinfrastructure funded by NSF under grant OCI-0959097 and an IBM Shared University Research (SUR) 
Award at Rice University, and 
the Extreme Science and Engineering
Discovery Environment (XSEDE) by NSF under Grant No.\ DMR170109.
Q.S. acknowledges the hospitality of the Aspen Center for Physics
(NSF under Grant No.\ PHY-1607611).

\par
\bibliography{2imp_SU2} 
\newpage

\onecolumngrid 
\section{Continuous-time Quantum Monte Carlo Method} 

In the continuous-time quantum Monte Carlo method, we perform a unitary transformation \cite{werner2007efficient,lang1962kinetic} to remove the $z$ component spin-boson coupling, using the generator ${ \cal S} = g(S_{1}^{z}-S_{2}^{z}) \sum_{p} \frac{1}{\omega_{p}} (\phi^{z\dagger}_{p}+\phi^{z}_{p}) $. The new Hamiltonian $\tilde{H} = e^{\cal S} He^{-\cal S}$ can be separated into two parts $\tilde{H}_0+\tilde{H}_1$, where
$\tilde{H}_{0}=\tilde{\epsilon}_{d} \tilde{d}_{i\sigma}^{\dagger} \tilde{d}_{i\sigma} + \sum_{i} \tilde{U} n_{i\uparrow} n_{i\downarrow} + \tilde{I} S_{1}^{z}S_{2}^{z}
+\sum_{{\bf k},\sigma} \epsilon_{\bf k} c_{{\bf k}\sigma}^{\dagger}c_{{\bf k}\sigma}
+ \sum_{p,\alpha} \omega_{p} {\phi^{\alpha}_{p}}^{\dagger} \phi^{\alpha}_{p} $, $\tilde H_{1}=\sum_{i}\left(H_{c,i}+ (-1)^{i+1} H_{b,i} \right) +H_{I}$ + h.c., with $H_{c,i}= \sum_{\bf{k},\sigma} V e^{i{\bf k \cdot r_{i} } } \tilde{d}_{i\sigma}^{\dagger} c_{{\bf k},\sigma} $, $H_{b,i}=g  \tilde{S}_{i}^{+} \Phi^{-} $, and $H_{I}=(I/2) \tilde{S}_{1}^{+}  \tilde{S}_{2}^{-}$. Here $\tilde{d}_{i,\sigma}=e^{\cal S} {d}_{i,\sigma} e^{-\cal S} =d_{i,\sigma} e^{-\frac{g}{2} s_{\sigma} p_{i} \sum_{p}\frac{1}{\omega_{p}} (\phi_{p}^{z\dagger} - \phi_{p}^{z} )}$ is the dressed fermionic operator, with $s_{\sigma}=\pm 1$ for $\sigma=\uparrow/\downarrow$ and $p_{i}=\pm 1$ for $i=1,2$. $\tilde{S}^{+}_{i}= \tilde{d}_{i\uparrow}^{\dagger} \tilde{d}_{i\downarrow}$, $\Phi^{-}=1/\sqrt{2} \sum_{p}\left( (\phi_{p}^{x \dagger} + \phi_{p}^{x}) - i ( \phi_{p}^{y \dagger} +  \phi_{p}^{y} )  \right)$. $\tilde{U}$, $\tilde{\epsilon_{d}}$ and $\tilde{I}$ are the renormalized coupling constant, given by $\tilde{U}=U+(g^{2}/2) \sum_{p} 1/\omega_{p}^{2}$, $\tilde{\epsilon_{d}}=\epsilon_{d}-(g^{2}/4) \sum_{p} 1/\omega_{p}^{2}\textbf{ }$ and $\tilde{I}=I+ 2g^{2}  \sum_{p} 1/\omega_{p}^{2}$.

Now, we are able to express the partition function ${\cal Z}$ in terms of an expansion of $\tilde{H}_{1}$ under the interaction representation of $\tilde{H}_{0}$,
\begin{eqnarray}
{\cal Z}&=&\sum_{n^{c}_{i},n^{b}_{i},\bar{n}^{c}_{i},\bar{n}^{b}_{i},n^{I},\bar{n}^{I} }  (-1)^{ p}  \times \text{Tr}[e^{-\beta \tilde{H}_{0}} {\cal T_{\tau}}
\nonumber \\
&\times& 
\frac{1}{n^{I}!}  \left( \int_{0}^{\beta} H_{I}(\tau) d\tau \right)^{n^{I}} \nonumber \
\frac{1}{\bar{n}^{I}!} \left( \int_{0}^{\beta} H^{\dagger}_{I}(\tau) d\tau \right)^{\bar{n}^{I}}
\nonumber \\
&\times& \prod_{i=1,2} 
\frac{1}{n^{c}_{i}!}  \left( \int_{0}^{\beta} H_{c,i}(\tau) d\tau \right)^{n_{i}^{c}} 
\frac{1}{\bar{n}^{c}_{i}!} \left( \int_{0}^{\beta} H^{\dagger}_{c,i}(\tau) d\tau \right)^{\bar{n}_{i}^{c}}\nonumber \\
&\times& \frac{1}{n^{b}_{i}!}  \left( \int_{0}^{\beta} H_{b,i}(\tau) d\tau \right)^{n_{i}^{b}}
\frac{1}{\bar{n}^{b}_{i}!} \left( \int_{0}^{\beta} H^{\dagger}_{b,i}(\tau) d\tau \right)^{\bar{n}_{i}^{b}}
 ] \, ,
\end{eqnarray}
where $p=n^{c}_{1}+\bar{n}^{c}_{1}+  n^{c}_{2}+\bar{n}^{c}_{2} + n_{1}^{b}+\bar{n}_{1}^{b}+n^{I}+\bar{n}^{I}$. Then
each term in the expansion is sampled by the Metropolis-Hastings algorithm \cite{metropolis1953equation,hastings1970monte}. 

The $(-1)^{p}$ factor in the expression could potentially cause a sign problem. However, it is always positive in this model. Notice that the number of conduction electrons at each site is a conserved quantity, meaning that $n_{i}^{c}=\bar{n}_{i}^{c}$. Besides, $S_{1}^{z}$, $S_{2}^{z}$, and number of bosons are all conserved quantity, such that $n_{1}^{b}+n^{I}=\bar{n}_{1}^{b}+\bar{n}^{I}$, $n_{2}^{b}+\bar{n}^{I}=\bar{n}_{2}^{b}+{n}^{I}$, and $n_{1}^{b}+n_{2}^{b}=\bar{n}_{1}^{b}+\bar{n}_{2}^{b}$. Under these conditions, $p$ is always even and thus $(-1)^{p}=1$. 


\section{Quantum phase transition} 
\begin{figure}[t]
\captionsetup[subfigure]{labelformat=empty}
  \centering
 	\mbox{\includegraphics[width=1\columnwidth]{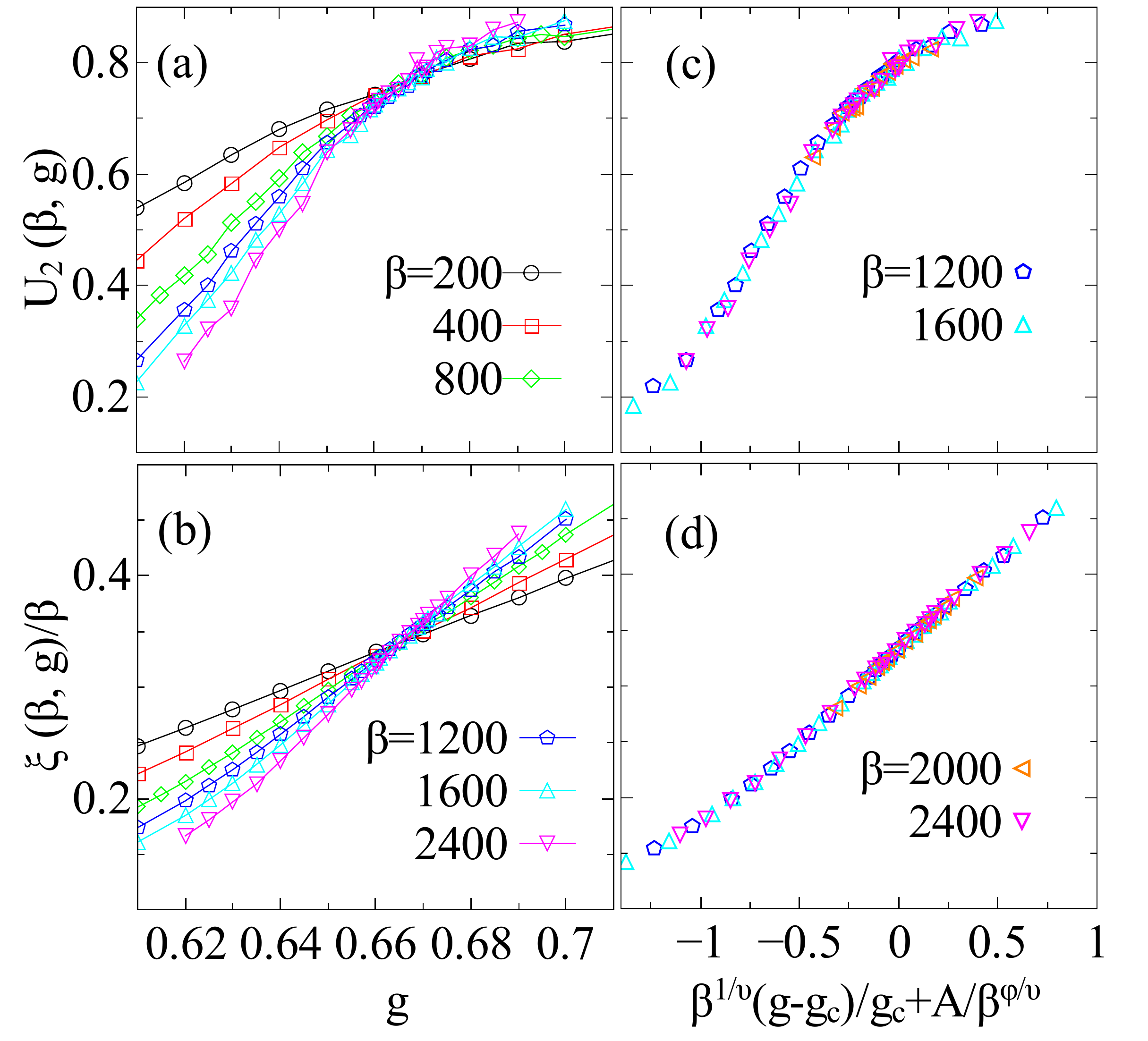}}
\caption
{Left: Binder cumulant $U_{2}$ (a) and the correlation length divided by inverse temperature $\xi/\beta$ (b) vs. bosonic coupling $g$ with $I=0.2$ at the labeled inverse temperature $1/\beta$. Both quantities gives a crossing at critical bosonic coupling $g_{c}=0.665(5)$. Right: Scaling collapse of $U_{2}$ (c) and $\xi$ (d) with a correlation length exponent $\nu^{-1}=0.34$, critical coupling $g_{c}=0.66$ (from $U_{2}$) and $\nu^{-1}=0.36$, $g_{c}=0.67$ (from $\xi/\beta$).}
\label{fig:2}
\end{figure}

To further establish the presence of a QCP between the Kondo screened phase and the LM phase, we calculate the Binder cumulant for a 3 components order parameter, 
$U_{2}=
5/2-{3{\langle  (\bf{m} \cdot \bf{m} )^{2} \rangle } }/{2{\langle  \bf{m} \cdot \bf{m} \rangle^{2}  } }$,
and correlation length $\xi$ of staggered spin correlation function $\chi_s^\alpha(\tau)$ \cite{binder1981finite,sandvik2010computational}.
Here ${\bf m}$ is the staggered magnetization with $m^{\alpha}=\langle \frac{1}{\beta} \int_{0}^{\beta} \left( S_{1} ^{\alpha}  (\tau) -S_{2} ^{\alpha}  (\tau) \right) d\tau \rangle$
and $\chi^{\alpha}_{s}(\tau) = \langle T_{\tau} ( S_{1} ^{\alpha}(\tau)  -S_{2} ^{\alpha} (\tau) )( S_{1} ^{\alpha} -S_{2} ^{\alpha}  ) \rangle $  \cite{cai2019bose}.

At the transition from Kondo screened to LM phase, both $U_{2}$ and $\xi/\beta$ should display a crossing at the critical coupling $g=g_{c}$, or in other words, become independent of $\beta$, the system size. As shown in Fig.\,\ref{fig:2} (a)(b), this is indeed what we have observed, where we plot $U_{2}$ and $\xi/\beta$ versus tuning parameter $g$ for various choices of $\beta$. In fact, the finite-size scaling hypothesis at a second-order QCP predicts that they should be described by the following scaling form near the QCP,
\begin{eqnarray}
U_{2}(g,\beta)&=&\tilde{U}_{2} \left( \beta^{1/\nu}(g-g_{c})/g_{c} + A/\beta^{\phi/\nu} \right) \, ,
\label{eq:U}\\
\xi(g,\beta) &=& \beta  \tilde{\xi} \left( \beta^{1/\nu}(g-g_{c})/g_{c} + A/\beta^{\phi/\nu} \right) \, .
\label{eq:xi}
\end{eqnarray}
$\tilde{U}_{2}$ and $\tilde{\xi}$ are universal functions, $g_{c}$ the critical coupling, $\nu$ the correlation length exponent. The term  $A/\beta^{\phi/\nu}$ captures the correction to scaling from the leading irrelevant operator. 
By performing scaling collapse analysis, we find the optimal choices for $g_{c}$ and $\nu$ are $g_{c}=0.66(1)$, $\nu^{-1}=0.34(4)$ (from $U_{2}$) and $g_{c}=0.67(1)$, $\nu^{-1}=0.36(4)$ (from $\xi$). From this we estimate the actual critical value to be $g_{c}=0.665(5)$ and the correlation length exponent of the Kondo destruction QCP to be $\nu^{-1}=0.35(5)$, which are consistent with the results derived from fidelity susceptibility $\chi_F$. The discrepancy is due to the larger scaling corrections of $\chi_F$.

\section{The procedure to calculate the pairing susceptibilities}
 
To measure the pairing susceptibility, we use the formula of the four-point correlation function shown in Ref.~\cite{gull2007performance}. However, due to the infinite separation in the SU(2) case, $\langle T_{\tau}  \Delta^{\dagger}_{\uparrow\downarrow} (\tau) \Delta_{\downarrow\uparrow}  \rangle$ and $\langle T_{\tau} \Delta^{\dagger}_{\downarrow\uparrow} (\tau) \Delta_{\uparrow\downarrow}  \rangle$ can not be measured via this formula, since they do not conserve the occupation number of $d_{i\sigma}$.
We can then utilize the SU(2) symmetry and obtain these pairing susceptibilities via:
$
\langle {\cal T_{\tau}}  \Delta^{\dagger}_{\uparrow\downarrow} (\tau) \Delta_{\downarrow\uparrow}  \rangle
=\langle {\cal T_{\tau}}  \Delta^{\dagger}_{\uparrow\uparrow} (\tau)\Delta_{\uparrow\uparrow} \rangle
-\langle {\cal T_{\tau}}  \Delta^{\dagger}_{\uparrow\downarrow} (\tau)\Delta_{\uparrow\downarrow} \rangle,
\langle {\cal T_{\tau}}  \Delta^{\dagger}_{\downarrow\uparrow} (\tau) \Delta_{\uparrow\downarrow}  \rangle
=\langle {\cal T_{\tau}}  \Delta^{\dagger}_{\downarrow\downarrow} (\tau)\Delta_{\downarrow\downarrow} \rangle
-\langle {\cal T_{\tau}}  \Delta^{\dagger}_{\downarrow\uparrow} (\tau)\Delta_{\downarrow\uparrow} \rangle
$ \cite{hoshino2016electronic}.

\section{Ising anisotropic model}
The Hamiltonian of Ising anisotropic BFAM is
\begin{eqnarray}
H&=&\sum_{i,\sigma} \epsilon_{d} d_{i\sigma}^{\dagger} d_{i\sigma} + \sum_{i} U n_{i\uparrow} n_{i\downarrow}  
+I S^{z}_{1} S^{z}_{2}
\nonumber \\
&+&\sum_{{\bf k},\sigma} \epsilon_{\bf k} c_{{\bf k}\sigma}^{\dagger}c_{{\bf k}\sigma}
+\sum_{i,{\bf k},\sigma} \left( V e^{i {\bf k \cdot r_{i}}} d_{i\sigma}^{\dagger} c_{{\bf k}\sigma} + h.c. \right) \nonumber \\ 
&+& \sum_{p} \omega_{p} {\phi^{z}_{p}}^{\dagger} \phi^{z}_{p} + g \sum_{p}( S_{1}^{z}-S_{2}^{z} )  ({ \phi_{p}^{z}}^{\dagger}
+\phi_{p}^{z}) \, .
\end{eqnarray} 
, where only $z$-component of RKKY interaction and spin-boson coupling are kept.
The effect of conduction electrons is described by two hybridization functions
\begin{eqnarray}
&&\Gamma(\epsilon) = \pi V^2 \sum_{\bf k} \delta(\epsilon-\epsilon_{\bf k}) \approx \Gamma_0\Theta(D-|\epsilon|) \nonumber  \\
&&\Gamma_{a}(\epsilon) = \pi V^2 \sum_{\bf k} \delta(\epsilon-\epsilon_{\bf k})\cos({\bf k}\cdot ({\bf r_1}-{\bf r_2}))\approx 
\gamma \epsilon \Gamma_0\Theta(D-|\epsilon|) \, ,
\end{eqnarray}
where $\Gamma(\epsilon)$($\Gamma_a(\epsilon)$) describes the effective hybridization from fermionic bath between two $d$ electrons of the same (different) sites \cite{finite_sep}. By tuning parameter $\gamma$, we are able to change the separation between two $d$ electrons: $\gamma=0$ denotes infinite separation $|{\bf r_1}-{\bf r_2}|=\infty$ and $\gamma=1$ represents $|{\bf r_1}-{\bf r_2}|=a$ with $a$ the lattice spacing. The finite separation generates an effective SU(2)-symmetric Heisenberg interactions of strength $2\gamma^2\Gamma^2/(\pi^2 U)$ between two $d$ electrons. Thus, we can estimate the effective spin-spin interactions $J_zS^z_1S^z_2 +J_p(S^x_1S^x_2+S^y_1S^y_2 )$ from fermionic bath, bosonic bath and RKKY interactions: $J_z = I_z+2g^2+2\gamma^2\Gamma^2/(\pi^2 U)$ and $J_p = 2\gamma^2\Gamma^2/(\pi^2 U)$. 
In the calculation, we take $\gamma=0.2$ and quantify the Ising anisotropy near QCP $J_p/J_z \approx 0.2 $. Other parameters are taken to be $U=-2\epsilon_d=0.001,\Gamma=0.25,s=0.8,D=1,\Lambda=1$

\end{document}